\documentclass[3p,times]{elsarticle}

\usepackage{ecrc}

\usepackage{epstopdf}
\usepackage{color}

\volume{00}

\firstpage{1}

\journalname{Nuclear Physics A}

\runauth{Berges et al.}

\jid{nupha}

\jnltitlelogo{Nuclear Physics A}

\usepackage{graphicx}
\usepackage{amsmath,amssymb}
\begin{document}

\begin{frontmatter}



\title{Turbulent thermalization process in high-energy heavy-ion collisions}

%% For multiple authors, replace the above by:

\author[label1,label2]{J\"urgen Berges}
\author[label3]{Bj\"orn Schenke}
\author[label3]{S\"oren Schlichting}
\author[label3]{Raju Venugopalan}

\address[label1]{Institut f\"ur Theoretische Physik, Universit\"at Heidelberg, Philosophenweg 16, 69120 Heidelberg, Germany}
\address[label2]{ExtreMe Matter Institute EMMI, GSI Helmholtzzentrum, Planckstraße 1, 64291 Darmstadt, Germany}
\address[label3]{Physics Department, Brookhaven National Laboratory, Upton, NY 11973, USA}

\begin{abstract}
We discuss the onset of the thermalization process in high-energy heavy-ion collisions from a weak coupling perspective, using classical-statistical real-time lattice simulations as a first principles tool to study the pre-equilibrium dynamics. Most remarkably, we find that the thermalization process is governed by a universal attractor, where the space-time evolution of the plasma becomes independent of the initial conditions and exhibits the self-similar dynamics characteristic of wave turbulence \cite{Berges:2013eia}. We discuss the consequences of our weak coupling results for the thermalization process in heavy-ion experiments and briefly comment on the use of weak coupling techniques at larger values of the coupling.  
\end{abstract}


\end{frontmatter}

% \linenumbers


\section{Introduction}
\label{Introduction}
The question of how exactly a thermalized Quark-Gluon Plasma (QGP) is formed in high-energy heavy-ion collisions is one of the major challenges in our current theoretical understanding of the experiments carried out at RHIC and the LHC. While practically all phenomenological models are based on the assumption that a close to thermal equilibrium state can be reached on a time scale of $\sim 1~{\rm fm}/c$, so far no clear theoretical understanding of this behavior has been established.

One of the major challenges in this regard is that the typical values of the coupling constant probed at present collider facilities $(\alpha_s\sim 0.3)$ are not necessarily in a suitable range for weak or strong coupling methods. Nevertheless significant progress in the understanding of the thermalization process has come from the study of these two limiting cases, where first principles calculations of the out-of-equilibrium dynamics are feasible \cite{Berges:2012ks}.

In this talk we discuss recent progress in the understanding of the thermalization process at weak coupling, which has been achieved by classical-statistical real-time lattice simulations \cite{Berges:2013eia}. Within their range of validity -- at weak coupling  $(\alpha_s \ll 1)$ and high occupancy $(f(t,p) \gg 1)$ \cite{Aarts:2001yn} -- these provide a first principles description of the non-equilibrium dynamics and allow for unprecedented insights into the thermalization process. We discuss the implications for the thermalization process at weak coupling in Sec.~\ref{WeakCoupling} and present an extrapolation towards more realistic values of the coupling. We also comment on the use of classical-statistical method at larger values of the coupling in Sec.~\ref{StrongCoupling} and present our conclusions in Sec.~\ref{Conclusions}. 

\vspace*{1cm}
\begin{figure}[t!]
\begin{center}
\includegraphics*[width=0.85\textwidth]{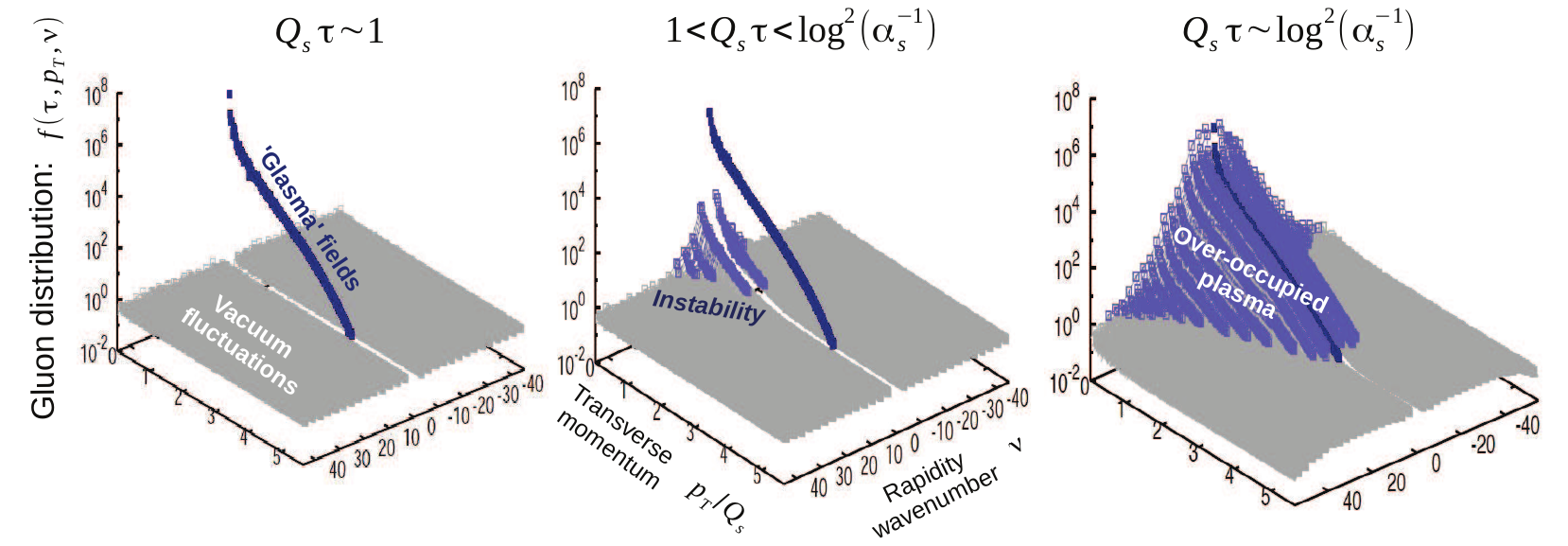}
\caption{Time evolution of the gluon distribution at early times $0 \lesssim Q\tau \lesssim \log^{2}(\alpha_{s}^{-1})$. While at early times the gluon distribution is dominated by the boost-invariant $(\nu=0)$ background, plasma instabilities lead to an over-occupied plasma on a time scale $Q\tau \sim \log^{2}(\alpha_{s}^{-1})$. Simulations are performed with (next-to-leading order) CGC initial conditions \cite{Epelbaum:2013waa} at very weak coupling ($\alpha_s\sim 10^{-6}$) for lattice size $256^2 \times 256$ and lattice spacings $Qa=1/2$ and $a_{\eta}=1/128$.}
\label{fig:EarlyTimes}
\end{center}
\end{figure}

\section{Thermalization process at weak-coupling}
\label{WeakCoupling}
\textit{Initial state.}
In the limit of weak-coupling and high collider energies the dynamics of the collision can be efficiently described in the color-glass condensate (CGC) framework \cite{Gelis:2010nm,Albacete:2014fwa}. Within this framework the initial state, formed immediately after the collision of heavy nuclei, is characterized by strong boost invariant color fields $(A_{\mu}^{a} \sim 1/g)$ \cite{Kovner:1995ts} usually referred to as the ``Glasma''. Even though the coupling constant is weak, the characteristic field amplitudes are large $\sim 1/g$ and the system is strongly interacting.\\

\textit{Early time dynamics.}
The boost invariant (2+1D) Yang-Mills dynamics of the ``Glasma'' has been studied in great detail \cite{Krasnitz:1998ns,Lappi:2006fp}. However, since the ``Glasma'' is a highly anisotropic state, it is unstable with respect to small vacuum fluctuations $(a_{\mu}^{a} \sim 1)$ which break the longitudinal boost invariance. While this was realized a long time ago \cite{Mrowczynski:1993qm}, and several studies have been performed since, the spectrum of vacuum fluctuations in the CGC framework was only obtained recently \cite{Epelbaum:2013waa}. We performed 3+1 D Yang-Mills simulations for these initial conditions to study the effect of instabilities at weak coupling ($\alpha_s\sim 10^{-6}$) where the classical-statistical framework is manifestly robust. Our preliminary results are summarized in Fig.~\ref{fig:EarlyTimes} showing the single particle gluon distribution at different times of the evolution. While initially the spectrum is dominated by the boost invariant ``Glasma'' fields we find that plasma instabilities lead to a (quasi-) exponential growth of low momentum modes \cite{Romatschke:2006nk}. As a result, an over-occupied plasma -- characterized by a non-perturbatively large occupancy of low momentum modes -- is formed on a short time scale which is parametrically given by $Q\tau \sim \log^{2}(\alpha_s^{-1})$  (c.f. \cite{Romatschke:2006nk}), where $Q$ denotes the characteristic momentum scale. We find that at this stage of the evolution the system is still quite anisotropic and far from equilibrium. Nevertheless, this suggests that in weak coupling, the initial conditions of [16] may in principle be represented by the initial conditions proposed in our prior work for times $Q\tau \gtrsim \log^2 (1/\alpha_S)$.\\

\textit{Onset of the thermalization process \& Classical attractor.}
The question of how the subsequent approach to thermal equilibrium occurs has been studied extensively in kinetic theory frameworks \cite{Baier:2000sb,Bodeker:2005nv}. However, since different thermalization scenarios have been developed based on parametric estimates, further progress relies on classical-statistical simulations which have the ability to distinguish which (if any) of the scenarios is realized. Since the numerical resources required to perform simulations for the entire evolution from $Q\tau=0^{+}$ up to very late times are very high, we instead employ a general parametrization of the over-occupied state at $Q\tau_{0}=\log^{2}(\alpha_s^{-1})$ -- with a variable occupancy $n_0/\alpha_s$ and variable momentum space anisotropy $\xi_0$ (see \cite{Berges:2013eia} for details) -- to study the subsequent onset of the thermalization process.

\begin{figure}[t!]
\begin{center}
\begin{minipage}[t]{0.45\linewidth}
\includegraphics*[width=\textwidth]{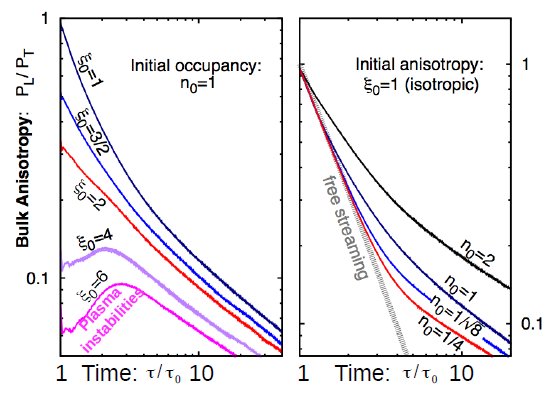}
\caption{ Time evolution of the bulk anisotropy (ratio of longitudinal to transverse pressure) for various initial conditions (see \cite{Berges:2013eia} for details). After a short transient regime different initial conditions all lead to the same universal scaling behavior, where the bulk anisotropy of the plasma increases as a function of time.}
\label{fig:Pressure}
\end{minipage}
\hspace{0.5cm}
\begin{minipage}[t]{0.45\linewidth}
\includegraphics*[width=\textwidth]{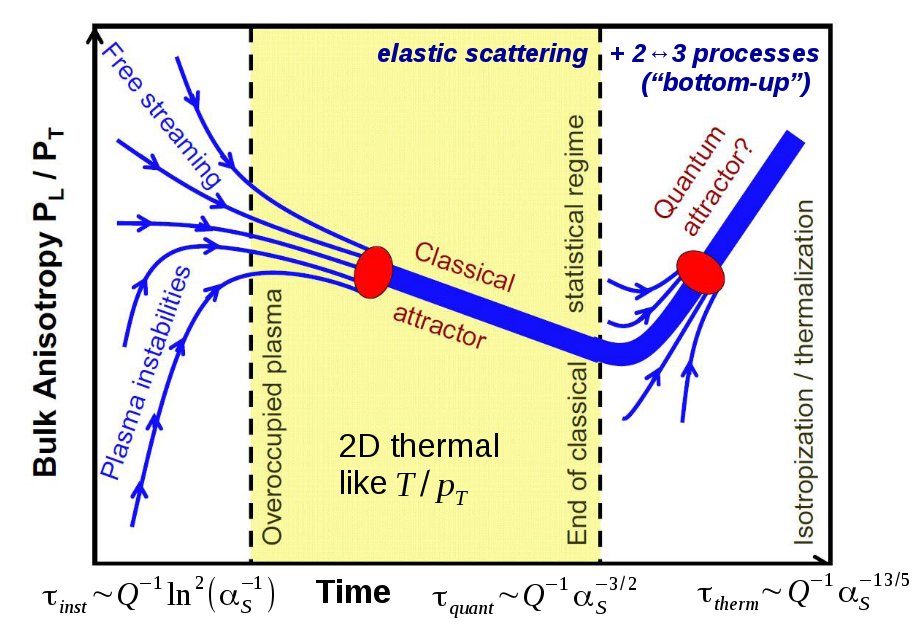}
\caption{ Illustration of the weak coupling thermalization process at the example of the time evolution of the bulk anisotropy. On a time scale $Q\tau \sim \log^{2}(\alpha_s^{-1})$ the evolution reaches a classical attractor solution where the anisotropy increases \cite{Berges:2013eia}. Beyond the classical regime $(Q\tau \gtrsim \alpha_s^{-3/2})$ inelastic processes are expected to lead to isotropization and equilibration of the plasma \cite{Baier:2000sb}.}
\label{fig:Cartoon}
\end{minipage}
\end{center}
\end{figure}

We find that on a time scale $\tau \sim \tau_{0}$ the evolution becomes insensitive to the details of the initial conditions and exhibits a universal scaling behavior, as shown at the example of the bulk anisotropy $P_L/P_T$ in Fig.~\ref{fig:Pressure}. A more detailed analysis reveals that the dynamics in the vicinity of this classical attractor becomes self-similar, and allows one to express the gluon distribution in terms of a time independent scaling function $f_S$ according to
\begin{eqnarray}
f(\tau,p_T,p_z)=\frac{(Q\tau)^{\alpha}}{\alpha_S} f_{S}\Big((Q\tau)^{\beta} p_T,(Q\tau)^{\gamma} p_z\Big)\;, 
\end{eqnarray}
at sufficiently late times \cite{Berges:2013eia}.  Such a behavior is characteristic of the phenomenon of wave turbulence and has been observed in a variety of far-from-equilibrium systems \cite{Micha:2004bv}. We find that the observed values of the scaling exponents $\alpha\simeq -2/3$, $\beta \simeq 0$ and $\gamma \simeq 1/3$  \cite{Berges:2013eia} are consistent with small angle elastic scattering as the dominant process and confirm the onset of the ``bottom-up'' thermalization scenario \cite{Baier:2000sb}. The competition between longitudinal momentum broadening via small-angle scattering and the red-shift due to the longitudinal expansion leads to a decrease of the typical longitudinal momenta as $p_z/Q \sim (Q\tau)^{-1/3}$,  while the typical transverse momenta remain approximately constant $p_T/Q\sim const$. This explains the observed increase of the bulk anisotropy in Fig.~\ref{fig:Pressure}. At the same time, the single particle occupancy decreases as $f(\tau,p_T\sim Q) \sim (Q\tau)^{-2/3}/\alpha_S$ and becomes of order unity on a time scale $Q\tau_{quant}\sim \alpha_S^{-3/2}$ when quantum effects can no longer be neglected. Beyond $\tau_{quant}$ the classical-statistical framework becomes inapplicable and one has to resort to a kinetic description.\\

\textit{Towards equilibrium.}
Following the bottom-up thermalization scenario \cite{Baier:2000sb}, which agrees with our lattice simulations within the respective range of validity, the thermalization process in the quantum regime ($Q\tau_{quant}\gtrsim \alpha_S^{-3/2}$) proceeds via inelastic processes which provide a road towards complete equilibration and isotropization on a time scale $Q\tau_{therm}\sim \alpha_S^{-13/5}$. When extrapolating these parametric estimates towards realistic values of the coupling constant at RHIC/LHC energies $(Q\sim 2~GeV,~\alpha_S\sim0.3)$, one obtains reasonable values for the end of the classical regime $\tau_{quant}\sim 0.6~{\rm fm}/c$ and the thermalization time $\tau_{therm}\sim 2~{\rm fm}/c$, which are in line with phenomenological assertions. We also note that more recently there has been significant progress in understanding the detailed dynamics in the quantum regime and refer the interested reader to \cite{Kurkela:2014tea}.

\section{Beyond weak coupling}
\label{StrongCoupling}
We also presented in our talk preliminary results for the time evolution of the energy-momentum tensor at larger values of the coupling constant $(g=0.15-0.3)$ following the work of Epelbaum and Gelis \cite{Gelis:2013rba}. We find that in this case the next-to-leading order (vacuum) contributions to the stress energy tensor can exceed the leading order (boost-invariant) contribution by orders of magnitude, such that a suitable renormalization procedure becomes inevitable. When we apply the vacuum subtraction method outlined in \cite{Gelis:2013rba}, a clear rise of the longitudinal pressure can be observed already at very early times $Q \tau=1-2$. Unfortunately, we find that the results obtained in this way show a strong  dependence on the lattice cut-off. We attribute this strong cut-off dependence to an unphysical decay of the vacuum, which is an artifact of the classical-statistical description when used beyond its range of validity \cite{Moore:2001zf}.

\section{Conclusions}
\label{Conclusions}
We studied the thermalization process in high-energy heavy-ion collisions from a weak coupling perspective, using classical-statistical real-time lattice simulations. Our findings are compactly summarized in Fig.~\ref{fig:Cartoon}, which illustrates the pre-equilibrium evolution at the example of the bulk anisotropy of the system. We find that on a time scale which is parametrically given by $Q\tau\sim \log^{2}(\alpha_S^{-1})$ the plasma reaches a classical attractor solution, where the evolution becomes insensitive to the details of the initial conditions and leads to an increase of the bulk anisotropy. The observed behavior is consistent with the first stage of the bottom-up thermalization scenario \cite{Baier:2000sb}, and can be understood as a competition between small angle elastic scattering processes and the longitudinal expansion of the system. Surprisingly, we find no indications that plasma instabilities play a significant role for the dynamics in this regime.

Beyond the time scale $Q\tau_{quant}\gtrsim \alpha_S^{-3/2}$, where the system has reached its maximal anisotropy, the occupancy of high momentum modes becomes small and a classical-statistical description is no longer applicable. Instead, following the bottom-up scenario, inelastic processes start to dominate and lead to equilibration and isotropization on a time scale $Q\tau_{therm}\gtrsim \alpha_S^{-13/5}$. When extrapolating the weak-coupling results towards realistic values of the coupling one obtains reasonable values for the thermalization time $\tau_{therm}\sim 2~{\rm fm}/c$. We thus conclude that weak-coupling thermalization is not in contradiction with the phenomenological observations at RHIC and LHC.

We also discussed the possibility to extend the use of weak-coupling techniques and perform real-time simulations directly at larger values of the coupling. We observe that present simulations at large couplings $g\geq 0.1$ are seriously affected by ultra-violet problems \cite{Moore:2001zf} which can strongly influence the interpretation of numerical results. Clearly, the question of how to extend weak-coupling methods towards more realistic values of the coupling constant at present collider energies remains an important topic for future studies.\\

\textit{Acknowledgments.} This research used resources of the National Energy Research Scientific Computing Center, which is supported by the Office of Science of the U.S. Department of Energy under Contract No. DE-AC02-05CH11231. This work was supported in part by the German Research Foundation (DFG). BPS,SS and RV are supported under DOE Contract No. DE-AC02-98CH10886.  SS gratefully acknowledges a Goldhaber Distinguished Fellowship from Brookhaven Science Associates. BPS is supported by a DOE Office of Science Early Career Award.

\end{document}